\documentclass[journal=jpclcd, manuscript=letter]{achemso}
\usepackage{graphicx}
\usepackage{booktabs}

\usepackage[normalem]{ulem}
\usepackage[dvipsnames]{xcolor}

\title{Peak Broadening in Photoelectron Spectroscopy of Amorphous Polymers: the Leading Role of the Electrostatic Landscape}

\author{Laura Galleni}
\affiliation[KU Leuven]
{Department of Chemistry, KU Leuven, Celestijnenlaan 200F, 3001 Leuven, Belgium}
\alsoaffiliation[IMEC]
{Imec, Kapeldreef 75, 3001 Leuven, Belgium}
\email{laura.galleni@imec.be}

\author{Arne Meulemans}
\affiliation[KU Leuven]
{Department of Chemistry, KU Leuven, Celestijnenlaan 200F, 3001 Leuven, Belgium}

\author{Faegheh S. Sajjadian}
\affiliation[KU Leuven]
{Department of Chemistry, KU Leuven, Celestijnenlaan 200F, 3001 Leuven, Belgium}
\alsoaffiliation[IMEC]
{Imec, Kapeldreef 75, 3001 Leuven, Belgium}

\author{Dhirendra P. Singh}
\affiliation[IMEC]{Imec, Kapeldreef 75, 3001 Leuven, Belgium}

\author{Shikhar Arvind}
\affiliation[KU Leuven]
{Department of Chemistry, KU Leuven, Celestijnenlaan 200F, 3001 Leuven, Belgium}
\alsoaffiliation[IMEC]
{Imec, Kapeldreef 75, 3001 Leuven, Belgium}

\author{Kevin M. Dorney}
\affiliation[IMEC]{Imec, Kapeldreef 75, 3001 Leuven, Belgium}

\author{Thierry Conard}
\affiliation[IMEC]{Imec, Kapeldreef 75, 3001 Leuven, Belgium}

\author{Gabriele D'Avino}
\affiliation[cnrs]{Grenoble Alpes University, CNRS, Grenoble INP, Institut Néel, 38042 Grenoble, France}

\author{Geoffrey Pourtois}
\affiliation[IMEC]{Imec, Kapeldreef 75, 3001 Leuven, Belgium}

\author{Daniel Escudero}
\affiliation[KU Leuven]
{Department of Chemistry, KU Leuven, Celestijnenlaan 200F, 3001 Leuven, Belgium}

\author{Michiel J. van Setten}
\affiliation[IMEC]{Imec, Kapeldreef 75, 3001 Leuven, Belgium}
\email{michiel.vansetten@imec.be}

\date{\today}

\begin{document}

\maketitle

\begin{tocentry}
\centering
\includegraphics[height=2in]{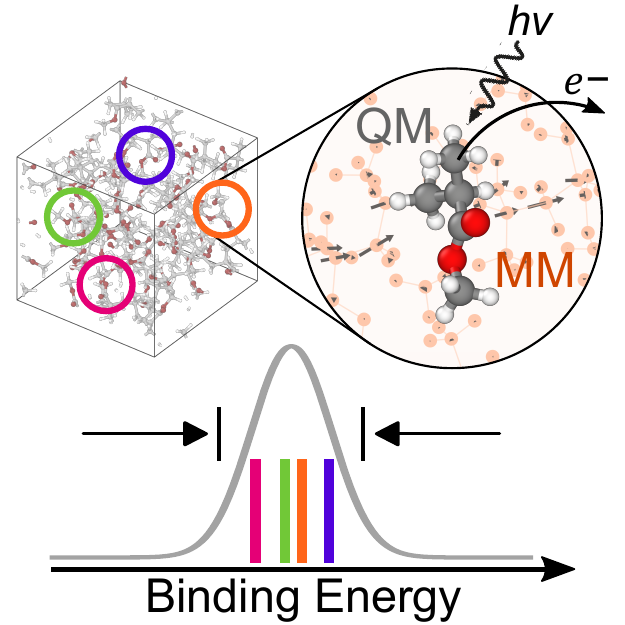}
\end{tocentry}

\begin{abstract}
    The broadening in photoelectron spectra of polymers can be attributed to several factors, such as light source spread, spectrometer resolution, finite lifetime of the hole state, and solid-state effects. Here, for the first time, we set up a computational protocol to assess the peak broadening induced for both core and valence levels by solid-state effects in four amorphous polymers by using a combination of density functional theory, many-body perturbation theory, and classical polarizable embedding. We show that intrinsic local inhomogeneities in the electrostatic environment induce a Gaussian broadening of $0.2$-$0.7$~eV in the binding energies of both core and semi-valence electrons, corresponding to a full width at half maximum (FWHM) of $0.5$-$1.7$~eV for the investigated systems. The induced broadening is larger in acrylate- than in styrene- based polymers, revealing the crucial role of polar groups in controlling the roughness of the electrostatic landscape in the solid matrix.
\end{abstract}

Photoelectron spectroscopy (PES) allows to study the composition of solids, liquids, and gaseous compounds by measuring the binding energies (BEs) of electrons in the material. In PES measurements, the electrons are extracted by the photoelectric effect using either x-rays (x-ray photoelectron spectroscopy, or XPS) or ultra-violet light (ultra-violet photoelectron spectroscopy, or UPS).\cite{suga2021photoelectron} XPS is mostly used to probe the BEs of core-electrons, whereas UPS is used for valence- to semi-core-electrons. The interpretation of experimental XPS and UPS spectra is often complicated by the relatively large peak broadening, which tends to lead to overlapping peaks. 

The peak broadening ($\Delta E$) in PES of polymers originates from two types of sources\cite{bagus2013interpretationXPS}: (i) \emph{extrinsic} sources, such as the light source spread and the spectrometer resolution ($\Delta E_{ext}$), and (ii) \emph{intrinsic} sources, such as the hole finite lifetime ($\Delta E_{\tau}$) and solid-state effects, such as disorder and inhomogeneities in the material ($\Delta E_{solid}$). Assuming uncorrelated sources of Gaussian broadening, the total width is the convolution of all contributions:

\begin{equation}
    \Delta E^2 = \Delta E_{ext}^2 + \Delta E_{\tau}^2 + \Delta E_{solid}^2
    \label{eq:deltaE}
\end{equation}

Typically, the fitted values for the full width at half maximum (FWHM) of experimental spectra are in the range of $1.2$-$1.6$~eV. The FWHM induced by extrinsic sources is generally in the range of $0.25$-$0.85$~eV\cite{assetsthermofisher, kratos}, while the lifetime of a photohole is of the order of femtoseconds, which corresponds to an additional broadening of about $0.1$-$0.2$~eV\cite{nicolas2012lifetime, carroll2000carbon}. 

Here, we propose a protocol based on first-principles calculations to estimate for the first time the contribution of solid-state effects to the broadening in XPS and UPS peaks of four common amorphous polymers: poly(methyl methacrylate) (PMMA), poly(t-butyl methacrylate) (PBMA), polystyrene (PS), and poly(4-hydroxystyrene) (PHS). These polymers were chosen as a test case for their vast range of applications\cite{serrano2020acrylate, scheirs2003polystyrenes} and in continuity with our previous work.\cite{galleni2022modeling}

An ideal model for theoretical PES spectra should yield the number of photoelectrons extracted from the sample as a function of their BEs. This requires three ingredients: (i) the density of states (DOS), i.e., the distribution probability of electronic states as a function of their BEs; (ii) the photoionization cross-section, which is the probability of extracting an electron from an orbital given its BE and photon energy; and (iii) a model for kinetic energy loss due to the scattering of photoelectrons in their way out of the sample, which leads to an increasing background for larger BEs. Secondary effects such as satellite peaks are neglected here. Although (ii) and (iii) may be important to reach a quantitative interpretation of the experimental spectra, most of the information is already contained in the DOS. This is especially true for XPS spectra, where the photoionization cross-section is practically independent of the BEs\cite{fadley1970electronic} and the background can be easily subtracted using known expressions.\cite{engelhard2020introductoryXPSbackground} Therefore, in this work, we neglect (ii) and (iii) and focus on the simulation of the DOS, by calculating the electron BEs.

Several theoretical methods are available to calculate BEs. A first approximation consists in taking the Hartree-Fock energies with opposite sign, as stated in Koopman’s theorem.\cite{Vines2018Prediction} In the literature, Kohn-Sham energies from density functional theory (DFT) are also often considered, although these values are not physical as they represent the single-particle energy of the non-interacting auxiliary system. Another method to calculate binding energies is $\Delta$SCF, where the BEs are calculated as the difference between the energy of the cation and that of the neutral system.\cite{Vines2018Prediction, Norman2018Simulating} Many-body perturbation theory in its many flavours, such as the $GW$ formalism,\cite{Aryasetiawan1998TheGWMethod, Onida2002, Golze2019Compendium, vanSetten2013gwImplementation} are also often used. Previous works showed that $GW$ yields very accurate valence and core-level BEs at an affordable computational cost.\cite{VanSetten2015GW100, Knight2016AccurateIpEaGW, VanSetten2018AssessingGWcore, Golze2020Accurate, li2022benchmark, mukatayev2022noblegas} 
However, the comparison with experiments requires introducing a peak broadening as a phenomenological parameter.

A popular technique to model amorphous solid materials is hybrid quantum mechanics/molecular mechanics (QM/MM).\cite{morzan2018QMMMspectroscopy} In this approach, the atomic system is divided into two regions: a QM fragment which is modeled with a high-level quantum-mechanical technique, and the MM environment, which is described with a lower-level classical model. The combination of high- and low-level theories provides a good compromise between accuracy and computational cost.
A typical QM/MM scheme for the simulation of PES is the so-called electrostatic embedding.\cite{morzan2018QMMMspectroscopy} In this scheme, the MM region is described with a classical polarizable model of atomistic resolution. Various works report BEs or ionization potentials calculated in a QM/MM framework with $\Delta$SCF\cite{niskanen2013hybrid, loytynoja2014quantumEthanolWater, ehlert2015combinedPVAwithQMMM, loytynoja2017quantumPMMA, ge2022qmmmN1s} or $GW$\cite{li2016combiningGWandPol,Li_DAvino_PRB18, tirimbo2020quantitative, tirimbo2020excited}. However, only few of these works focus on amorphous solids\cite{ehlert2015combinedPVAwithQMMM, tirimbo2020quantitative}, and although some attempts have been made on polymer chains\cite{loytynoja2017quantumPMMA, tirimbo2020excited}, only the work by Ehlert et al. reports calculated binding energies of a full amorphous model of a polymer\cite{ehlert2015combinedPVAwithQMMM}, although calculated only for core-levels using $\Delta$SCF. 

\begin{figure}
    \centering
    \includegraphics[width=6.5in]{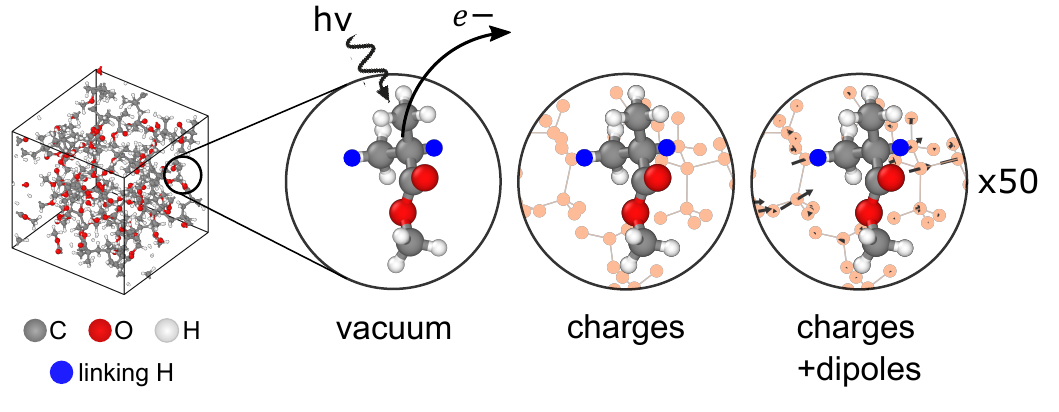}
    \caption{Scheme of the three approximations considered in this work: hydrogen-terminated repeat unit in vacuum, embedded in point charges, and embedded in point charges and dipoles. Calculations are repeated for $50$ different repeat units of the polymer matrix.}
    \label{fig:embedding}
\end{figure}

In this work, we improve on the results reported in the literature by (i) investigating four new polymer systems, (ii) considering amorphous atomistic models optimized at a higher level of theory, namely a combination of force-fields and DFT, and (iii) calculating both valence and core level binding energies using GW in the full-analytical approach\cite{vanSetten2013gwImplementation}, which does not require the explicit creation of a hole, as opposed to $\Delta$SCF, while allowing to treat core-electrons, in contrast to other implementations of GW such as contour-deformation or the plasmon-pole model used in similar works\cite{li2016combiningGWandPol,Li_DAvino_PRB18, tirimbo2020quantitative, tirimbo2020excited}. The electrostatic embedding scheme used in this work is illustrated in Figure~\ref{fig:embedding}.

To generate the atomic coordinates of each polymer structure, we follow a multi-step protocol: first, a polymer of 50 repeat units is generated by means of our own polymer builder code, imposing a nominal density within $0.1$~g~cm$^{-3}$ from the experimental value (more details in the Supporting Information, SI); then, the structure is optimized by sequentially applying time-stamped force-bias Monte Carlo (TFMC)\cite{mees2012uniformTFMC} and Broyden–Fletcher–Goldfarb–Shanno (BFGS) optimization\cite{BROYDEN1970} both using DFT forces and imposing periodic boundary conditions as implemented in the CP2K software package.\cite{Khne2020} 

Each one of the $50$ repeat units of the polymer structure is then terminated with hydrogens and considered as the QM region for a separate QM/MM calculation to obtain the binding energies. The hydrogen-terminated repeat unit is henceforth called fragment. This choice of fragments has been justified for these particular polymers in a previous work\cite{galleni2022modeling} and is expected to be reasonable for most non-conjugated polymers where there is no significant delocalization of the electronic states. The BEs of each fragment are computed with single-shot $GW$ ($G_0W_0$) in the fully-analytical approach\cite{vanSetten2013gwImplementation} on top of DFT at the BH-LYP\cite{becke1988densityb3-lyp1, lee1988developmentb3-lyp2, becke1993newBH-LYP}/def2-TZVPP level of theory in combination with resolution of identity\cite{eichkorn1995auxiliary, Weigend2005, weigend2006accurate}. 
The MM atomistic environment corresponds to a cloud of about 200000 atomic charges and dipoles surrounding the QM region, obtained by replicating the simulation box in the three dimensions within a sphere of radius 80~\AA, as to converge electrostatic potential. Fractional charges and induced dipoles in the neutral ground state are obtained from self-consistent microelectrostatic calculations (MESCal code)\cite{davino2014mescal} for the neutral ground state of the entire simulation box, imposing periodic boundary conditions.
MM embedding does not consider dynamical polarization effects arising from the reaction of the dielectric environment to the photo-hole, that, however, negligibly contribute to energetic disorder in bulk systems.\cite{DAvino_2016}
The intrinsic solid-state broadening is then estimated from the standard deviation of the BEs over the $50$ fragments.

\begin{figure}
    \centering
    \includegraphics[width=6.5in]{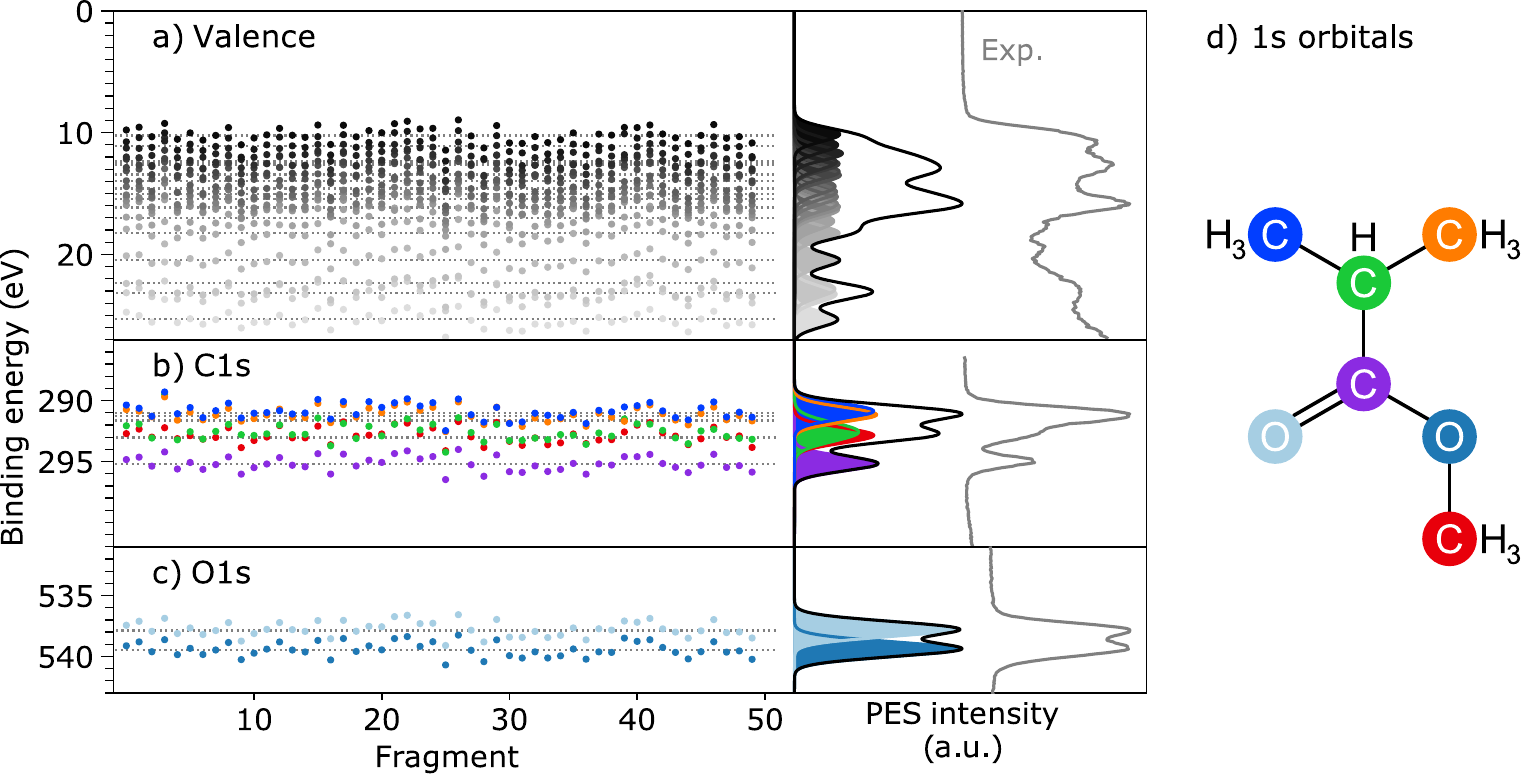}
    \caption{Distribution of the calculated binding energies for (a) valence, (b) C 1s, and (c) O 1s orbitals of $50$ fragments of PMMA embedded in charges+dipoles, and (d) scheme of the corresponding C 1s and O 1s orbitals. Dotted lines in the left panel are the calculated values for the fragment relaxed in vacuum. Experimental XPS and UPS spectra of the polymer are normalized and aligned to the theory; the background is not subtracted.}
    \label{fig:pmma_emb_dip}
\end{figure}

Figure \ref{fig:pmma_emb_dip} shows the BEs of the O 1s, C 1s and the upper $21$ valence levels of $50$ fragments of PMMA embedded in charges+dipoles. Similar figures for the other polymers are reported in the SI, Figures~S1-S12. 
These Figures reveal significant fluctuations in the calculated BEs for both core and valence levels, ultimately resulting in a sizeable peak broadening. 
Such an \emph{energetic disorder} is a direct consequence of the roughness of the electrostatic landscape arising from the differences in the local environment of the $50$ fragments, as confirmed by the correlations between BEs and the electrostatic potential, reported in the SI, Figure~S13. 
The distribution of BEs appears symmetric around the average values, which in turn do not differ significantly from the values calculated on the isolated fragment relaxed in vacuum (dotted lines, see SI, Table~S2).
To simulate the spectra, we plot a Gaussian peak with width equal to the standard deviation over the $50$ BEs, centered on the average values. The area of each Gaussian peak is normalized as we neglect here the contribution of the photo-ionization cross-section and assume that each orbital contributes equally to the photoelectron spectra. To justify the choice of a Gaussian line-shape, we perform a Shapiro-Wilk test\cite{shapiro1965analysis} for each BE and find that only one peak out of 27 deviates from a normal distribution (see SI, Figure~S16).
The resulting theoretical DOS obtained from the sum of these Gaussian peaks is in qualitative agreement with the experiment. Notice that no background was subtracted from the experimental data, which is why a discrepancy is visible at higher BEs. Moreover, a rigid shift to the calculated spectrum was applied  to align the most intense peak to the experimental one, allowing a direct comparison of the relative peak positions. The shift, which is different for core and valence BEs and is in the range of $3$-$7$~eV, accounts for several factors, such as 
the approximations involved in the computational scheme (incomplete basis set, DFT functional, $GW$ self-energy), the different reference energies used in theory and experiment (vacuum and Fermi edge, respectively), and charging effects, as discussed in Ref.~\cite{galleni2022modeling}.
Interestingly, the spread in energies is similar for both valence and core levels. 
This indicates that the electrostatic landscape varies over a length scale that is comparable to the molecular size, as confirmed by the analysis of the potential correlation function (see SI, Figure~S15).

\begin{figure}
    \centering
    \includegraphics[width=6.5in]{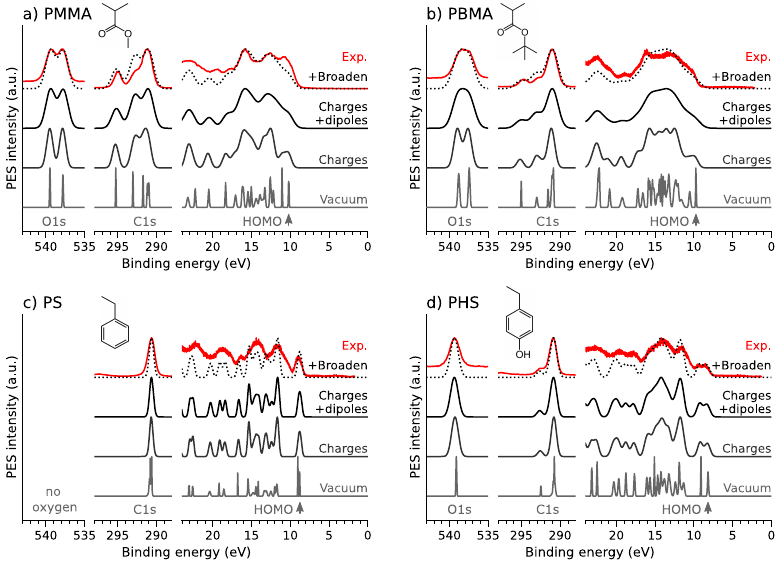}
    \caption{Experimental and calculated PES spectra for O 1s, C 1s, and valence regions of a) PMMA, b) PBMA, c) PS, and d) PHS in the three considered approximations and with the addition of a broadening of $0.3$~eV to account for the finite lifetime and the instrumental resolution. Spectra are normalized and experimental data are aligned to the most intense theoretical peak in the charges+dipoles approximation, except for the UPS spectra of PBMA where the peak at $22.4$~eV is considered instead.}
    \label{fig:all_mol_all_emb}
\end{figure}

To ascertain that the spread is due to the local electrostatic environment and not to structural variations of the repeat units within the polymer, the calculation is repeated on each of the $50$ fragments in vacuum without point charges and dipoles, as shown in Fig.~\ref{fig:all_mol_all_emb}a, bottom spectrum. Interestingly, the broadening induced by local geometrical variations is negligible with respect to the electrostatic effects, for both core and semi-valence orbitals. This result is in agreement with our previous work on 1s core level BEs\cite{galleni2022modeling} and is not surprising, as core-electron BEs are weakly sensitive to the position of the neighboring atoms due to the strong localization of the 1s orbitals. However, it is remarkable that the effect of geometry is negligible also for valence orbitals, where the stronger delocalization should, in principle, lead to a larger effect on the BEs.
Moreover, the comparison between the charges and charges+dipoles embedding schemes show that the dipoles clearly contribute to the disorder, although their effect is smaller than that of charges alone.

\begin{table}[]
    \centering
    \caption{Average FWHM in eV calculated with $G_0W_0$ or DFT (parenthesis) over all O 1s, C 1s, and semi-valence BEs of $50$ fragments.}
    \begin{tabular}{lllll}
    \toprule
$\Delta E_{solid}$ (eV) &           PS &          PHS &         PMMA &         PBMA \\
\midrule
Vacuum              &  0.15 (0.09) &  0.20 (0.13) &  0.18 (0.15) &  0.20 (0.19) \\
Charges             &  0.44 (0.39) &  0.78 (0.74) &  0.96 (0.96) &  0.98 (0.97) \\
Charges + dipoles &  0.49 (0.43) &  0.91 (0.86) &  1.48 (1.47) &  1.68 (1.66) \\
    \bottomrule
    \end{tabular}    
    \label{tab:mean_sigma}
\end{table}

To investigate the generality of these conclusions, the calculations were repeated on three additional polymers. The results are reported in Fig.~\ref{fig:all_mol_all_emb}b-d and Table \ref{tab:mean_sigma}. In all cases, the impact of structural variations is negligible, whereas the presence of charges and dipoles induces a significant contribution to the peak broadening. Interestingly, the electrostatic spread depends on the type of polymer, being minimum in PS ($0.49$~eV) and maximum in PBMA ($1.68$~eV). 
Our calculations reveal a good correlation between the spread in BEs and the magnitude of the fragment dipole moments, see SI, Figure~S14. 
This suggests that the energetic disorder is mostly originating from the dipole fields of the randomly oriented repeat units, hence being larger in systems with polar groups, such as hydroxyl and acrylate ones.

To verify that the calculated values for the solid-state broadening in polymers are compatible with the presence of additional sources of broadening, we add a FWHM of $0.3$~eV to the theoretical spectra simulated in the charges+dipoles approximation to account for the lifetime broadening and the extrinsic sources, as shown in Fig.~\ref{fig:all_mol_all_emb}. The total broadening is calculated with Eq.~\ref{eq:deltaE}, assuming $\Delta E_{\tau}=0.2$~eV and $\Delta E_{ext}=0.25$~eV which is the typical instrumental resolution for a monochromatized Al K$\alpha$ light source\cite{kratos}. The resulting spectra are reported in Fig.~\ref{fig:all_mol_all_emb} and show an excellent agreement with experiments. It should be noted that our calculations describe the static component of the energetic disorder, neglecting the dynamical contribution associated with nuclear thermal motion. Therefore, the very good agreement with experiments suggests that the static component dominates over the dynamical one, as observed in other amorphous systems\cite{DAvino_2016}. Nevertheless, the dynamical contribution may explain the small discrepancy observed in PS and PHS, where the simulated broadening is smaller than the experimental one.

Finally, we notice that the spread due to solid-state effects can be already retrieved at the DFT level (Table \ref{tab:mean_sigma}), showing that the effect of local electrostatic fluctuations is already captured by the Kohn-Sham energies. This result is not trivial and could not be assumed \emph{a priori}. Earlier work on the valence levels of molecular crystals showed that the electrostatic potential of the environment directly affects Kohn-Sham levels, having a minor effect on the GW exchange-correlation contribution.\cite{Li_DAvino_PRB18} The present study provides solid evidence that the same applies also to highly localized core levels in the strongly-varying potential of an amorphous system. In the case only peak broadening is of interest, one could consider reducing the computational cost by sampling energy levels over different fragments with DFT calculations in combination with electrostatic embedding. However, if one is interested in the band shapes over 5-10 eV energy span, as in the present study, quality of a GW calculation is required in order to accurately capture relative peak positions (chemical shifts), for which correlations effects beyond Kohn-Sham energies can be important.\cite{Mukatayev2023}

In conclusion, we have shown that a significant amount of broadening in photoelectron spectra of amorphous polymers arises from the effect of inhomogeneities in the local electrostatic landscape. First-principles calculations using a combination of many-body perturbation theory, density functional theory, and electrostatic embedding, show that the electrostatic variations alone lead to a FWHM of $0.5$-$1.7$ eV, and that larger contributions are expected in polymers with acrylate groups (PMMA and PBMA) than in those containing only a phenol (PHS) or a phenyl group (PS). On the contrary, the broadening induced by local variations in the polymer geometry are negligible. The computational protocol proposed here can be potentially applied to other disordered materials to provide insights into the physics of the photoelectric processes.

\begin{acknowledgement}
The authors acknowledge the contribution of Ilse Hoflijk (Imec) for performing the XPS measurements on PMMA. The authors acknowledge funding from the Imec Industrial Affiliation Program (IIAP). G.D. acknowledges support from the French ``Agence Nationale de la Recherche", project RAPTORS (ANR-21-CE24-0004-01).
\end{acknowledgement}

\section{Supporting information}
Computational and experimental details; distribution of calculated BEs in vacuum, charges, and charges+dipoles for PMMA, PBMA, PHS, and PS; mean absolute deviation of the average BEs with respect to the BEs of the fragments in vacuum; fluctuations of the atomic potential, dipole of the repeat unit, correlation of the atomic potential; p-values of Shapiro tests.

\bibliography{bibliography}

\end{document}